# Helical Magnetic Field in a Massive Protostellar Jet

A. Rodríguez-Kamenetzky[1], A. Pasetto[2], C. Carrasco-González[2], L. F. Rodríguez[2], J. L. Gómez[3], G. Anglada[3], J. M. Torrelles[4,5], N. R. C. Gomes[5], S. Vig[6], and J. Martí[7]
[1] Instituto de Astronomía Teórica y Experimental (IATE), Universidad Nacional de Córdoba (UNC), Córdoba, Argentina; adriana.rodriguez@unc.edu.ar
[2] Instituto de Radioastronomía y Astrofísica (IRyA), Universidad Nacional Autónoma de México (UNAM), Morelia, Michoacán, México
[3] Instituto de Astrofísica de Andalucía (IAA), Consejo Superior de Investigaciones Científicas (CSIC), Glorieta de la Astronomía s/n, E-18008 Granada, Spain
[4] Institut de Ciències de l'Espai (ICE-CSIC), Bellaterra, Barcelona, Spain
[5] Institut d'Estudis Espacials de Catalunya (IEEC), Barcelona, Spain
[6] Department of Earth and Space Science, Indian Institute of Space Science and Technology, Thiruvananthapuram 695 547, Kerala, India
[7] Departamento de Física, Escuela Politécnica Superior de Jaén, Universidad de Jaén, Campus Las Lagunillas s/n, 23071 Jaén, Spain
Received 2024 October 04; revised 2024 November 17; accepted 2024 November 29; published 2025 January 7

## Abstract

Highly collimated outflows (jets) are observed across a wide range of astrophysical systems involving the accretion of material onto central objects, from supermassive black holes in active galaxies to proto-brown dwarfs and stellar-mass black holes. Despite the diversity of their driving sources, it is believed that all jets are different manifestations of a single universal phenomenon. However, a unified explanation for their ejection and collimation remains elusive. In this study we present the first rotation measure analysis of the polarized synchrotron emission ever performed in a protostellar radio jet, which allows us to reveal its true 3D magnetic structure. Unlike extragalactic radio jets, which often exhibit faint counterjets, protostellar radio jets allow both the jet and the counterjet to be analyzed. This exceptional circumstance allows us to unveil the magnetic field structure of both components. Our findings provide the first solid evidence for a helical magnetic field within a protostellar jet, supporting the universality of the jet collimation mechanism.

*Unified Astronomy Thesaurus concepts:* Protostars (1302); Radio jets (1347); Interstellar magnetic fields (845); Interstellar synchrotron emission (856)

## 1. Introduction

The most widely accepted model to explain the launching and collimation of jets is based on magnetocentrifugal forces within the accretion disk, which cause ionized material in the inner disk to follow magnetic field lines. In this scenario, the disk's rotation generates a large-scale helical magnetic field configuration that effectively channels the outflowing material into a focused jet (e.g., R. D. Blandford & D. G. Payne 1982; R. E. Pudritz & C. A. Norman 1983; F. Shu et al. 1994; R. E. Pudritz & T. P. Ray 2019). Consequently, magnetic confinement shapes and collimates the material at a certain distance from the central object. Other mechanisms such as the "Poynting–Robertson (PR) cosmic battery" have also been proposed for the generation and amplification of magnetic fields in active galactic nuclei (AGNs; e.g., I. Contopoulos & D. Kazanas 1998; D. M. Christodoulou et al. 2008). In this mechanism, as the footpoints of the initially poloidal magnetic field lines are dragged by the rotating disk plasma, an overall helical magnetic field is generated.

A comprehensive understanding of jets emanating from astrophysical bodies necessitates mapping their magnetic field configuration, which is a challenging task. In young stellar objects (YSOs), one way to study the magnetic fields in protostellar molecular outflows and jets is through the measurement of the Goldreich–Kylafis effect (P. Goldreich & N. D. Kylafis 1981; S. Deguchi & W. D. Watson 1984; J. M. Girart et al. 1999; T. C. Ching et al. 2016; C.-F. Lee et al. 2017; P. C. Cortés et al. 2021). Furthermore, the magnetic field strength in protostellar jets can also be inferred from the analysis of shock-excited line ratios (e.g., P. Hartigan & A. Wright 2015). On the other hand, in AGN jets, the dominant emission mechanism is synchrotron radiation; thus, by analyzing the polarization pattern of the radio waves emitted by the jet, it is possible to determine the orientation of the magnetic field in different regions of the outflow. However, polarized emission can be affected by *Faraday* effects when passing through a magnetoionic medium, such as Faraday rotation, Faraday depolarization, and Faraday repolarization. There are several models available in the literature that describe the different depolarization/repolarization scenarios (i.e., B. J. Burn 1966; P. C. Tribble 1991; D. D. Sokoloff et al. 1998; F. Mantovani et al. 2009). By applying these models to observational data, in conjunction with rotation measure (RM) analysis, it is possible to infer the intrinsic polarization structure of a jet (e.g., A. Pasetto 2021). These techniques have been applied to the study of relativistic jets from AGNs, mainly close (up to a few parsecs) to the supermassive black hole, (e.g., J. L. Gómez et al. 2008; T. Hovatta et al. 2012; M. Mahmud et al. 2013; D. C. Gabuzda et al. 2014, 2015, 2017; TEHT Collaboration et al. 2021a, 2021b). Interestingly, the strongest evidence for a helical magnetic field has been found in the jet of the active galaxy M87, at kiloparsec scales from the central black hole (A. Pasetto et al. 2021), a remarkable result that suggests interactions between magnetic fields and matter at such large scales.

In the case of protostellar jets, radio emission is usually dominated by the free–free radiation of slow-moving, thermal electrons, resulting in nonpolarized radiation. However, recent studies have shown that a population of relativistic electrons can be present in protostellar jets that experience strong shocks to the medium (A. Rodríguez-Kamenetzky et al. 2016, 2017, 2019). To date, several protostellar jets exhibiting signs of nonthermal emission have been identified through the detection of negative spectral indices at centimeter wavelengths (e.g., L. F. Rodríguez et al. 1989; J. Martí et al. 1993; J. M. Girart et al. 2002; R. E. Ainsworth et al. 2014; A. Rodríguez-Kamenetzky et al. 2016, 2017; M. Osorio et al. 2017; S. Vig et al. 2018; W. O. Obonyo et al. 2019; V. Rosero et al. 2019; A. Sanna et al. 2019; A. G. Cheriyan et al. 2023). Nevertheless, polarized emission in these systems is very weak, having been detected and studied only in a single object (C. Carrasco-González et al. 2010). This is the case of the HH 80-81 jet, a particularly fast jet ($\sim 10^3$ km s$^{-1}$) driven by a massive protostar, IRAS 18162-2048 (J. Martí et al. 1993). The bolometric luminosity of the YSO is estimated to be between 10,000 and 14,000 times that of the Sun, assuming a distance of between 1.2 and 1.4 kpc (N. Añez-López et al. 2020). Furthermore, N. Añez-López et al. (2020) also conducted dynamic estimations of the mass and modeling of the system, inferring a stellar mass of approximately $\sim 20\,M_\odot$. Although polarization in this object was detected through sensitive observations using the classic Very Large Array (VLA), the detection was limited to a single wavelength (6 cm) with a narrow bandwidth (100 MHz). This prevented an analysis of the RM to eliminate Faraday rotation effects. Consequently, while this study confirmed the presence of relativistic particles in a protostellar jet and opened the possibility of investigating magnetic fields in these objects, these observations were insufficient to infer the 3D configuration of the magnetic field.

In this work we study the deepest polarization observations at centimeter wavelengths of the HH 80-81 jet. We perform the RM analysis for the first time in a protostellar jet, which allows us to reveal the true 3D structure of its magnetic field.

## 2. Observations and Methodology

The source IRAS 18162-2048, was observed with the VLA of the National Radio Astronomy Observatory (NRAO[8]) for 20 hr at the C band in C configuration, in full polarization mode (project 18B-290; PI: Carlos Carrasco-González). Observations were carried out in four sessions of 5 hr each in 2018: November 25 and December 4, 8, and 11, including observations of a standard flux/bandpass/polarization angle calibrator (3C286), as well as a phase calibrator (J1911−2006), and a leakage calibrator (J2355+4950). We used the standard frequency setup for continuum observations, i.e., 2 MHz width channels covering the entire frequency range (4–8 GHz). The phase center was R.A. (J2000) $=18^h19^m12\overset{s}{.}101$, decl. (J2000) $= 20°47'30\overset{''}{.}90$. Data calibration was done with the Common Astronomy Software Applications (CASA) package (version 6.1.2.7; J. P. McMullin et al. 2007) and the NRAO pipeline for VLA continuum observations, which was modified to include polarization calibration after the complex gain calibration. With this purpose we followed a procedure similar to that used in A. Pasetto et al. (2018). We first set the polarization calibrator model using the task SETJY in manual mode. This procedure was adopted because CASA models do not currently include polarization. Therefore, we supplied the flux density, spectral index, and polarization parameters (polarization fraction and angle) of 3C286. These parameters were obtained from polynomial fitting to the polarization properties of 3C286 and reference flux densities listed on the NRAO website. The corresponding Stokes I values were calculated using the Perley and Butler scale (R. A. Perley & B. J. Butler 2017). Cross-handed (RL, LR) delays per spectral window were solved using 3C286. The Leakage terms were corrected using the J2355+4950 calibrator, assumed to be an unpolarized source (CASA guides), which was observed at the end of each session. After calibrating for instrumental polarization, we calibrated the R–L phase to accurately obtain the polarization position angle. To determine the absolute electrical vector polarization angle, we use source 3C286, whose position angle is known and previously set with SETJY. The pipeline was run separately for each of the four observational sessions. Once calibrated, all data were concatenated into a single file for subsequent imaging and self-calibration.

The initial self-calibrated image revealed bright sources within the field of $13' \times 20'$, one of them highly variable, which proved challenging to clean using the TCLEAN task and elevated the overall image rms noise. To mitigate this, we subtracted these sources in the uv-plane by modeling their emission per observational scan and spectral window, as appropriate, employing the UVSUB task. After calibration and source subtraction, we proceed with the imaging of the jet. We construct images of the Stokes parameters I, Q, and U, restricting the data to the range 2–20 k$\lambda$ (>2 k$\lambda$ to avoid lobe contamination due to short baselines, and <20 k$\lambda$ to attempt to reproduce the observations of C. Carrasco-González et al. 2010). We use the TCLEAN task with natural weighting, and given the large bandwidth, we use multiscale multifrequency synthesis (see U. Rau & T. J. Cornwell 2011), setting the parameter nterms = 2 for Stokes I to account for spectral variation within the band. For Stokes Q and U we use nterms = 1. Polarized emission from the jet is imaged using task IMMATH in the poli and pola modes to obtain the polarized intensity (P) and polarization angle (A) maps, respectively.

### 2.1. Rotation Measure Analysis

When linearly polarized radiation passes through a magnetoionic medium, the polarization plane may experience a rotation. This phenomenon, known as the Faraday rotation effect, can be attributed to the fact that a linearly polarized wave can be decomposed into two opposite-handed circularly polarized components, each of which experiences a different phase velocity within the material. Consequently, a relative phase shift between the circularly polarized components is induced, resulting in the rotation of the polarization plane of the linearly polarized wave. In most of the astrophysical settings the rotation angle $\Delta\chi$ can be expressed as $\Delta\chi = \phi(n_e, B_\parallel)\lambda^2$, where $\lambda$ is the observed wavelength, $\chi$ is the polarization angle, and $\phi$ is the Faraday depth, which depends on the electron density in the plasma, $n_e$, and the magnetic field in the line of sight, $B_\parallel$. The general expression for the Faraday depth is written as (e.g., A. Pasetto 2021) $\phi[\text{rad/m}^2] = 8.1 \times 10^5 \int_L n_e B_\parallel dL$, where the electron density is expressed in cm$^{-3}$, the magnetic field in Gauss, and the path length $L$ in pc. Faraday rotation can cause the polarization angle and the fractional polarization (polarization degree) of the radiation to be altered. Modeling variations in the fractional polarization (depolarization/repolarization) is a complex process that requires extensive data across a wide range of wavelengths to accurately constrain model parameters. However, in the simplest case, i.e., in the presence of a uniform magnetic field in the medium (Faraday screen), the Faraday depth is identical to its polarized emission-weighted mean value, referred to as RM. The latter is therefore determined by fitting the linear equation $\chi(\lambda^2) = \chi_0 + \text{RM}\lambda^2$, where $\chi$ is the observed polarization angle and $\chi_0$ is the intrinsic polarization angle of the source. This equation is valid only when the RM fitting is restricted to regions of the $\lambda^2$ space where the polarization fraction is constant (e.g., M. Simard-Normandin et al. 1981; A. Pasetto 2021).

---

[8] The NRAO is a facility of the National Science Foundation operated under cooperative agreement by Associated Universities, Inc.

In this work, our deep observations allow us to study polarization properties with wavelength within the observed band, in the range ∼4–8 cm; thus, only the simplest model (Faraday screen) can be applied to study the polarized emission in this source. The complex representation of the polarized signal in the presence of a Faraday rotating screen is (B. J. Burn 1966)

$$p(\lambda) = p_0 e^{2i(\chi_0 + \text{RM}\lambda^2)}, \quad (1)$$

where $p_0$ and $\chi_0$ are the intrinsic fractional polarization and the intrinsic polarization angle (i.e., polarization properties of the source not affected by Faraday effects), respectively. These quantities are defined as $p = \frac{P}{I} = \sqrt{q^2 + u^2}$ and $\chi = \frac{1}{2}\arctan\frac{u}{q}$, where $q = Q/I$ and $u = U/I$ are the fractional values of the Stokes parameters $Q$ and $U$, as used by D. Farnsworth et al. (2011). This equation describes a constant dependence of $p$ with $\lambda^2$ and a linear behavior of the observed polarization angle $\chi = \chi_0 + \text{RM}\lambda^2$. Thus, in this simplified case, the polarization angle undergoes a rotation, but the polarization degree remains unchanged.

To proceed with the analysis of the rotation measure, we create images of the Stokes parameters $I$, $Q$, and $U$ for each of the ∼128 MHz width spectral windows (spw) across the total 4 GHz bandwidth. Thus, using the task TCLEAN we obtained 31 natural-weighted images for each of the Stokes parameters $I$, $Q$, and $U$, using hogbom deconvolver (an adapted version of Högbom Clean; J. A. Högbom 1974) and setting parameter $n$terms = 1. All images were convolved to the same beam size of $12'' \times 7''\!.5$ (PA = 7°). Next, we constructed a data cube by assembling the Stokes $I$ spw images. The cube contains 31 channel images of $2'\!.2 \times 3'\!.5$ size, each corresponding to the total emission $I$ at the central frequency $\nu$ of each spectral window, $I(\nu)$. The Stokes $Q$ and $U$ images underwent the same process. These cubes enable a pixel-by-pixel study of how the Stokes parameters and derived parameters vary with frequency.

In order to determine the polarization properties of the source, we worked with the data cubes of the Stokes parameters $I$, $Q$, and $U$ to remove the effects of Faraday rotation from the polarized emission on a pixel-by-pixel basis. We proceeded by simultaneously fitting the fractional Stokes parameters $q(\lambda^2)$ and $u(\lambda^2)$ pixel by pixel, using the simplest equation of a one-component RM model (Equation (1)). This process results in three maps showing the spatial distribution of the RM, the intrinsic polarization angle ($\chi_0$), and the intrinsic fractional polarization ($p_0$). Therefore, we construct the magnetic field map by rotating polarization vectors through a 90° angle. The magnetic field vector map is then transformed into a streamline image by using the line integral convolution (LIC) technique (B. Cabral & L. C. Leedom 1993).

## 3. Results and Discussion

The observational results before performing RM analysis are those presented in Figure 1. In this figure we show the continuum emission of the entire extension of the jet over the full bandwidth (left), where the jet's driving source (IRAS 1862−2048) is labeled along with the famous Herbig–Haro (HH) objects HH 80, and HH 81 and the obscured HH object HH 81 N first studied at radio frequencies (e.g., J. Martí et al. 1993; A. Rodríguez-Kamenetzky et al. 2019), and recently also detected at optical and near-IR wavelengths (J. Bally & B. Reipurth 2023). These HH objects trace intense shocks of the jet against the ambient medium far from the driving protostar. Polarized emission was detected at smaller scales (∼0.4 pc from the protostar) and at several wavelengths, ranging from 3.75 to 7.5 cm, in both the jet (northeast lobe) and the counterjet (southwest lobe); see Figure 1 (center). Polarization vectors (white segments), indicating the direction of the electric field of the observed linearly polarized emission, are displayed over the total intensity ($I$) map (Figure 1, center); a color-coded map of the polarization degree (fractional polarization, $p$) of the polarized radiation ($P$) is also shown. The polarization degree decreases toward the jet's axis and increases toward the edges, consistent with the expected effects of Faraday rotation and depolarization (M. Lyutikov et al. 2005). Therefore, the results obtained for both the polarization vectors and the polarization degree map are fully consistent with those reported in Figure 1 of C. Carrasco-González et al. (2010). As an example of the analysis of polarization properties with wavelength conducted for each pixel in the image, we illustrate in the same figure (on the right) the variation of the fractional polarization $p$ and polarization angle $\chi$ as a function of wavelength squared at the emission peak of both the jet and the counterjet. It can be observed that both parameters exhibit variations with wavelength. This, in conjunction with the detection of a gradient in the degree of polarization across the jet's width, suggests that, most likely, the observed polarized emission originates in the jet and that it is affected by Faraday rotation effects (see, e.g., A. Pasetto 2021). It thus follows that the polarization vectors (white segments) would not trace the intrinsic direction of the magnetic field. Consequently, it is essential to perform an RM analysis to ascertain the intrinsic polarization vectors and, in turn, the true orientation of the magnetic field in the plane of the sky.

Now, in order to interpret the results after performing RM analysis let us consider the case of a jet with its axis in the plane of the sky, whose width can be angularly resolved. In the presence of a helical magnetic field we would expect to find a strong depolarization at the jet's central axis compared to its edges (M. Lyutikov et al. 2005). This depolarization occurs because the central region integrates radiation from various depths within the jet, each experiencing its own unique combination of magnetic field configuration and material properties. Additionally, we would also expect a transverse Faraday RM gradient due to the systematic change in magnetic field in the line of sight across the jet (e.g., R. A. Perley et al. 1984; M. Lyutikov et al. 2005; I. Contopoulos et al. 2009). A negative value of the RM indicates a magnetic field entering in the plane of the sky, while a positive value indicates a magnetic field approaching the observer. If the jet is inclined with respect to the plane of the sky, however, the RM gradient resulting from a helical magnetic field may comprise RM values of a single sign, depending upon the pitch angle of the helical field and the angle at which the jet is viewed (e.g., D. Gabuzda [2019](#)).

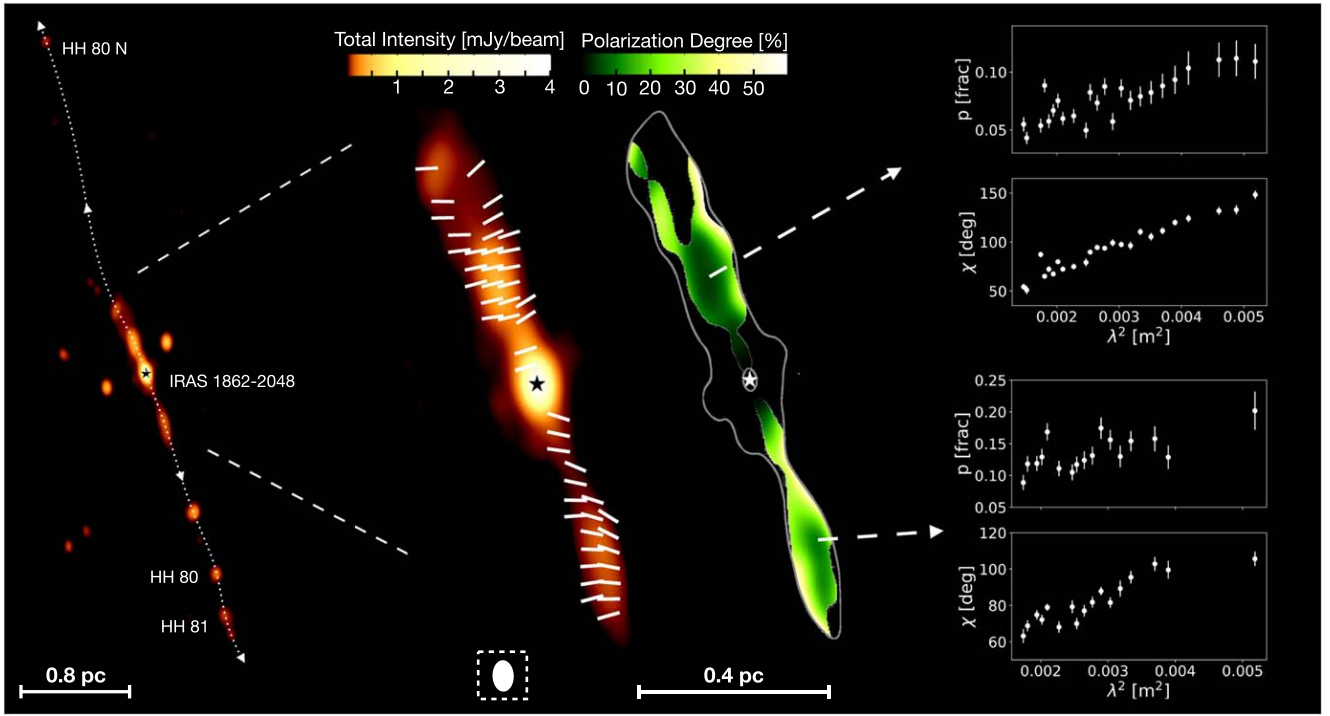

**Figure 1.** The HH 80-81 jet at 5 cm. We show natural-weighted images of the total intensity $I$ and polarization degree $p$ in color scale and polarization angle (vectors) with an angular resolution of $10'' \times 6''$, PA = 4°. The location of the driving source of the jet, IRAS 1862-2048, is indicated by a star symbol. Left: total intensity map of the entire radio jet, with dotted arrows indicating its precessional path. Dashed lines delineate the studied region, magnified in the center and right panels. Herbig–Haro objects, marking strong jet–ambient medium shocks, are also labeled. Center: polarization vectors (white segments) over total intensity (color scale); polarization degree of the emission (color scale) and total intensity (gray contours, corresponding to [10, 1000] × the rms of the I image). Both, the polarization vectors and polarization degree are shown over a total intensity signal-to-noise (SNI) > 10 and a polarized emission signal-to-noise (SNP) > 4, where the rms = 3 $\mu$Jy beam$^{-1}$. The synthesized beam is shown at the bottom left. The physical scale of 0.4 pc corresponds to 60'. Right: we show the variation of the polarization degree and polarization angle as a function of wavelength squared at the emission peak of both the jet and the counterjet.

Our deep observations allowed us to study polarized emission in narrow sub-bands (~128 MHz wide) with a sensitivity of 10–20 $\mu$Jy beam$^{-1}$. This enabled a detailed multiwavelength analysis of polarization properties across the entire observed bandwidth. Figure 2 illustrates that the model adequately describes the observed polarization properties within the observed range. The Stokes parameters $q$ and $u$ are accurately fitted with small residual errors, and the linear behavior of the polarization angle with wavelength is effectively reproduced. On the other hand, our results (Figures 1 and 2) suggest a slight increase of the polarization degree with $\lambda^2$. However, as previously noted, Equation (1) describes a constant dependence of the polarization fraction with $\lambda^2$. Consequently, any deviations from this model are not taken into account. As aforementioned, modeling depolarization/repolarization is a complex issue that requires extensive data across a wide range of wavelengths, which is beyond the scope of this study. It is noteworthy, however, that the simple Faraday screen model is capable of accurately recovering the behavior of the polarization angle, which is essential for determining the intrinsic magnetic field of the jet.

Figure 3 presents the results of the RM analysis, displayed above a specific threshold (see the figure caption). In the figure (left and center) we show streamline maps of the true magnetic field component in the plane of the sky, overlaid on the total intensity image and RM sign map, respectively, in color scale.

Moreover, in the same figure (on the right) we present a sketch illustrating the 3D structure of the magnetic field (indicated by yellow arrowed lines), as derived from the polarization analysis. The achieved angular resolution allowed for resolving the jet's width, thereby enabling the clear detection of gradients of RM (Figure 3, center) and polarization degree (previously shown in Figure 1, center) across the width of both the jet (NL) and the counterjet (SL). The RM map indicates the direction of the magnetic field along the line of sight. Blue regions indicate a magnetic field emerging from the plane of the sky, while regions in red indicate magnetic field lines in the opposite direction, pointing away from the observer. Additionally, the magnetic field component in the line of sight ($B_\parallel$) can be derived from the expression given for the Faraday depth in Section 2.1, assuming that $B_\parallel$ and the electron density ($n_e$) are constant within the jet: $B_\parallel[G] = (8.1 \times 10^5)^{-1} \frac{RM}{\text{rad/m}^2} \left(\frac{n_e}{\text{cm}^{-3}}\right)^{-1} \left(\frac{L}{\text{pc}}\right)^{-1}$. With this purpose, we first estimate $n_e$ and the jet's depth, $L$, at the distance where the synchrotron-emitting north lobe is located. Since the region of interest falls between ~20'' and ~50'' from the driving source of the jet, we take the mean value, $z = 35''$ corresponding to 0.24 pc, to perform calculations. Following A. Rodríguez-Kamenetzky et al. ([2017](#)), from Equation (14), and estimating from optical observations (J. Bally & B. Reipurth [2023](#)) the jet's half-width, $r_{HH} \sim 2.''5$, at the distance $z_{HH} \sim 300''$ where HH objects are located, it is possible to estimate the transversal radius of the jet, $r$, at the distance $z$. We obtain $r_{z=0.24\text{pc}} \simeq 0.''53$ corresponding to 0.0036 pc. Thus, the jet's cross section at the distance $z$ is $A_{z=0.24\text{pc}} = \pi r^2_{z=0.24\text{pc}} \simeq 3.87 \times 10^{32}$ cm$^2$. Next, following

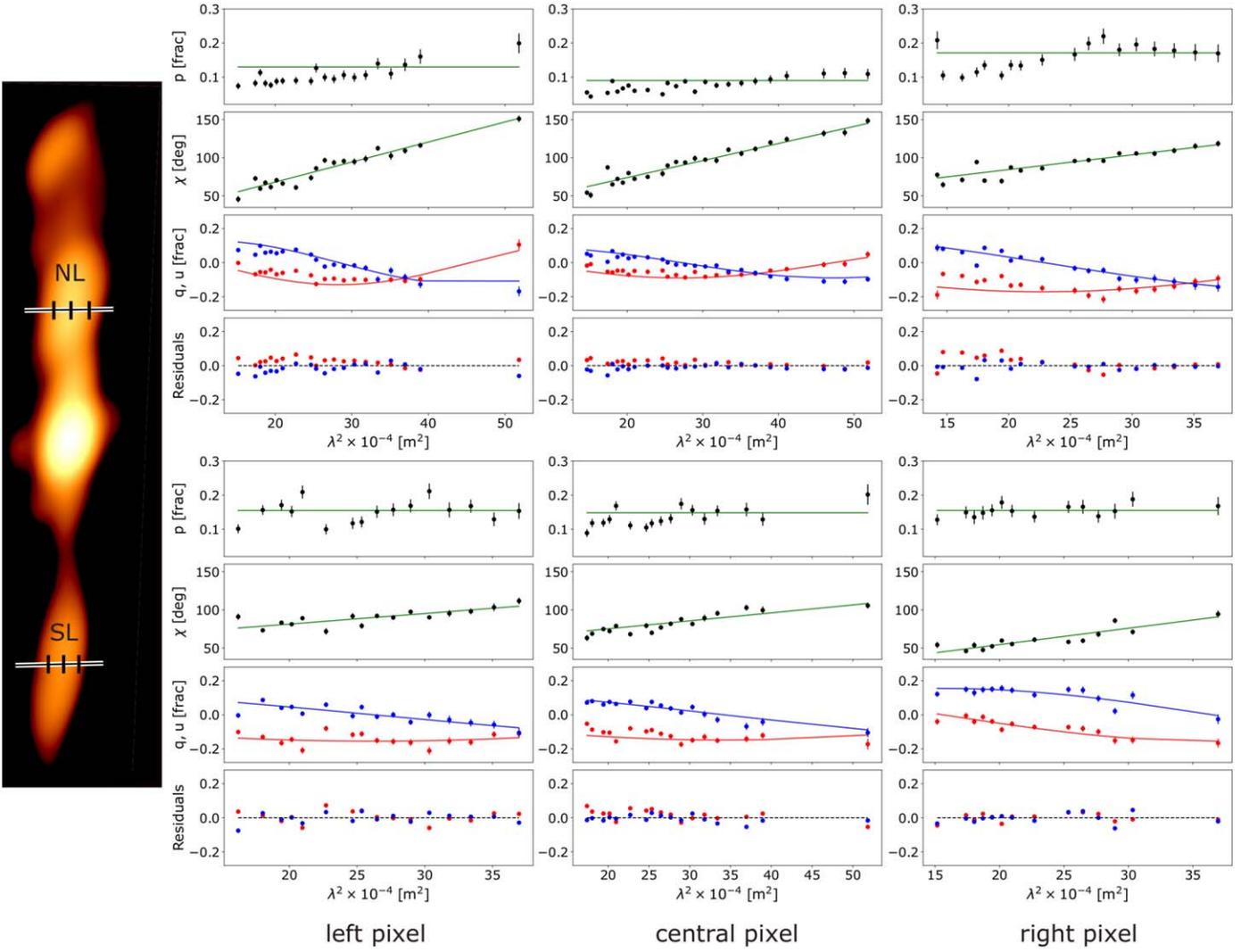

**Figure 2.** Comparison of the variation of polarization parameters ($p$, $\chi$, $q$ in red, and $u$ in blue) as a function of wavelength squared, for the simple Faraday screen model (solid line) and the observations (dots). Residuals of $q$ and $u$ with respect to the model predictions are also shown. The Faraday screen model fitting was conducted pixel by pixel. Six representative pixels are shown along a cross section of the jet axis, intersecting the emission peaks of both the jet (NL) and counterjet (SL), with the central point corresponding to each peak (see the illustrative image on the left; angular resolution and intensity scale can be found in Figure 1, center). The top and bottom panels show the behavior of polarization parameters with the wavelength squared for pixels selected in the north and south lobes, respectively. Error bars are indicated when larger than the symbols.

Equation (12) in A. Rodríguez-Kamenetzky et al. (2017), we estimate $n_e \sim 300\,\mathrm{cm^{-3}}$ (assuming a mass-loss rate of $\dot{M} = 6 \times 10^{-7} M_\odot\,\mathrm{yr^{-1}}$ and a jet velocity $v = 1000\,\mathrm{km\,s^{-1}}$). On the other hand, from the RM analysis, we derive RM values in the range 200–500 rad m$^{-2}$, with 350 rad m$^{-2}$ as the mean value, so that subtracting the Galactic RM (+160 rad m$^{-2}$ S. Hutschenreuter et al. 2022) results in 190 rad m$^{-2}$. Finally, assuming cylindrical symmetry we can consider $L = 2r_{z=0.24\mathrm{pc}} \simeq 0.0072$ pc, and therefore, with the estimated density and the mean value of the RM, an approximation for the magnetic field in the line of sight is obtained: $B_\parallel \sim 0.1$ mG. However, the process of determining the electron density is subject to uncertainties, and therefore this value should be considered as an order-of-magnitude estimate. It can be noticed that the strength of the magnetic field in the line of sight is consistent with the strength of the magnetic field in the plane of the sky derived by energy equipartition by C. Carrasco-González et al. (2010), namely, 0.2 mG. Therefore, the average magnetic field strength in the region would be of similar magnitude, i.e.,

$B = \sqrt{B_\parallel^2 + B_\perp^2} \sim 0.22$ mG. Furthermore, the inclination of the magnetic field with respect to the line of sight can be estimated as $i = \tan^{-1}(B_\perp/B_\parallel) \sim 60°$, differing only $\sim 10°$ with the jet inclination derived by N. Añez-López et al. (2020), i.e., 49°.

On the other hand, the observed RM gradients in the jet and counterjet exhibit clear opposition. Given the established counterclockwise rotation of the accretion disk (C. Carrasco-González et al. 2012; M. Fernández-López et al. 2023), these transverse gradients are indicative of a toroidal magnetic field, $B_T$, pointing in the direction of disk rotation in the northern hemisphere of the accretion disk and exhibiting an opposite orientation in the southern hemisphere. These results suggest that the twisting direction of the $B_T$-field lines follows the disk's rotation, whereas the field orientation is independent of the rotational motion. The orientation of the magnetic fields may depend on the mechanism generating them. For example, the PR cosmic battery that has been proposed to operate in AGNs predicts different magnetic field orientations in the jet

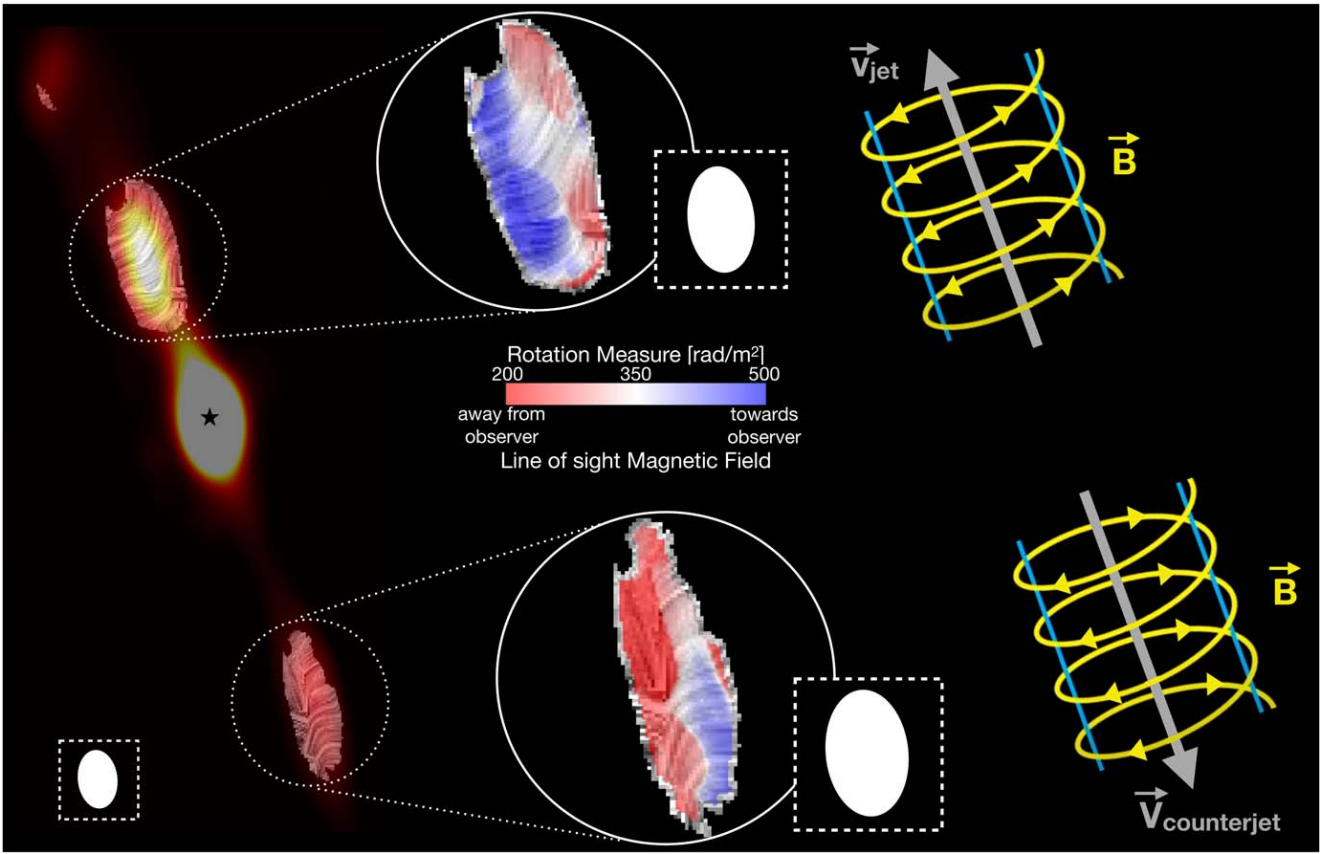

**Figure 3.** Results of the rotation measure (RM) analysis in the HH 80-81 jet. The left image shows the streamline image (obtained through the line integral convolution technique) of the component of the magnetic field parallel to the plane of the sky, over the total intensity map. In the middle panel we show the streamline image of the magnetic field parallel to the plane of the sky, over the values of the Faraday rotation obtained from our modeling of the Stokes parameters $q(\lambda^2)$ and $u(\lambda^2)$. The color scale of the RM indicates the direction of the magnetic field along the line of sight, i.e., red, away from the observer, and blue, toward the observer. Both, the magnetic field streamlines and the RM maps are shown for pixels with a signal-to-noise ratio (S/N) > 7 in both, the total intensity and the polarized emission. The $12'' \times 7''\!\!.5$ beam (PA = 7°) is shown as a white ellipse in both panels. The right panel shows a scheme depicting the 3D configuration of the magnetic field, exhibiting a helical topology, as suggested by our analysis of the polarization properties.

and counterjet (I. Contopoulos & D. Kazanas 1998; I. Contopoulos et al. 2006, 2009).

The analysis conducted provides strong evidence for the presence of a helical magnetic field within the HH 80-81 jet: (1) as previously discussed, the observed strong depolarization at the jet axis in comparison to its edges and the presence of transverse Faraday RM gradients across the jet are consistent with a helical magnetic field configuration (M. Lyutikov et al. 2005); (2) observing these properties in the polarized emission across the jet width, in addition to the variations observed in both polarization angle and polarization degree with wavelength, suggests the presence of internal depolarization. This implies that the observed polarized emission originates from the jet itself rather than the surrounding medium, thus tracing the jet's intrinsic magnetic field (see, e.g., A. Pasetto et al. 2021). Given that a helical configuration of the magnetic field is the result of a poloidal component originating in the protostar–disk system and a toroidal component generated by differential disk rotation, we can interpret our results in terms of poloidal and toroidal components projected in the plane of the sky. Figure 3 (left and center) reveals that the magnetic field streamlines are primarily toroidal, with a shift toward a poloidal component at its edges, i.e., in the center the toroidal component is dominant, while at the edges the poloidal component is more pronounced. Since the observed magnetic field is averaged along the line of sight, the dominant component detected may be affected by the jet's inclination (45°; N. Añez-López et al. 2020) and the pitch angle of the magnetic field (M. Lyutikov et al. 2005). On the other hand, theoretical models indicate that for the launching and collimation of a jet, a helical magnetic field is necessary in the vicinity of the protostar. At significant distances from it, the poloidal component is expected to decrease, yet collimation is possible due to the toroidal component. Thus, in this scenario, and considering a distance of ∼0.4 pc from the protostar (where the jet lobes are observed), two potential outcomes emerge. Either both components (poloidal and toroidal) are present at a distance of 0.4 pc from the protostar, and one dominates over the other due to the combination of the jet inclination and the pitch angle, or only the toroidal component exists at 0.4 pc (the poloidal component being very weak). Given that the toroidal component originates from the poloidal component by the rotation of the disk, the last case also implies a helical configuration of the magnetic field. Consequently, we provide conclusive evidence of the presence of a helical magnetic field intrinsic to the jet.

This is the first time that an analysis of the RM has been performed in a protostellar jet, allowing us to reveal the true 3D configuration of its magnetic field. Our findings provide the first solid evidence of a helical magnetic field in a protostellar jet. This in turn strongly supports the scenario of a universal mechanism for explaining all astrophysical jets, being in full agreement with C. Carrasco-González et al. (2010). Moreover, following the line of argumentation started in C. Carrasco-González et al. (2010), we underline the importance of studying radio jets in YSOs having the advantage of exhibiting thermal and nonthermal emission, allowing us to obtain physical parameters (such as electron density, among others), which are not easily obtained in AGN jets (see also the reviews of G. Anglada et al. 2015; G. Anglada et al. 2018). In this work we also show that the study of the magnetic field has advantages in YSO radio jets, as we can always observe both the approaching jet and the receding counterjet, whereas in AGN radio jets the counterjet is usually faint and only the approaching jet is clearly observed as a result of the strong Doppler boosting. This is also evidence of a helical magnetic field intrinsic to the disk–jet system and not a field from the surrounding medium.

The protostellar jet of HH 80-81 is powered by a relatively massive protostar and exhibits particularly intense radio emission. Nevertheless, the study of the magnetic field in this jet has only been possible after a major observational effort. Such studies should be extended to other objects, in particular those powered by solar-mass stars, where in some cases, techniques distinct from those described here have enabled the detection of intense magnetic fields that play an important role in the operation of their jets (e.g., see T. P. Ray et al. 1997; C.-F. Lee et al. 2017). Nevertheless, the reconstruction of the magnetic field in a large sample of these objects is beyond the capabilities of current instrumentation. The situation is expected to evolve in the near future with the advent of new, highly sensitive radio instrumentation such as the Square Kilometre Array (SKA[9]; see G. Anglada et al. 2015) and the Next Generation Very Large Array (ngVLA[10]), which will also allow polarization to be detected at smaller scales, closer to the protostars. Furthermore, the possibility of conducting similar RM studies of the Galactic center microquasars 1E 1740.7-2942 and GRS 1758-258 could provide insights into magnetic fields in these objects. In this context, we emphasize the significance of this study in laying the groundwork for future research aimed at understanding the role of the magnetic fields in the Universe.

## Acknowledgments

A.R.-K. acknowledges support from Consejo Nacional de Investigaciones Científicas y Técnicas (CONICET), and support from the European Research Executive Agency HORIZON-MSCA-2021-SE-01 Research and Innovation program under the Marie Skłodowska-Curie grant agreement number 101086388 (LACEGAL). G.A. is supported by grants PID2020-114461GB-I00, PID2023-146295NB-I00, and CEX2021-001131-S, funded by MCIN/AEI /10.13039/501100011033. N.R.C.G. acknowledges financial support from the European Research Council (ERC) under the European Union's Horizon Europe program (ERC Advanced grant SPOTLESS; No. 101140786). J.M.T. acknowledges partial support from the PID2020-117710GB-I00 grant funded by MCIN/AEI/10.13039/501100011033, and by the program Unidad de Excelencia María de Maeztu CEX2020-001058-M. J.M. acknowledges support by grant PID2022-136828NB-C42 funded by the Spanish MCIN/AEI/ 10.13039/501100011033.

*Facility:* VLA (NRAO)

*Software:* Common Astronomy Software Applications, CASA package (version 6.1.2.7; J. P. McMullin et al. 2007).

## ORCID iDs

A. Rodríguez-Kamenetzky https://orcid.org/0000-0002-4731-4934
A. Pasetto https://orcid.org/0000-0003-1933-4636
C. Carrasco-González https://orcid.org/0000-0003-2862-5363
L. F. Rodríguez https://orcid.org/0000-0003-2737-5681
J. L. Gómez https://orcid.org/0000-0003-4190-7613
G. Anglada https://orcid.org/0000-0002-7506-5429
J. M. Torrelles https://orcid.org/0000-0002-6896-6085
N. R. C. Gomes https://orcid.org/0000-0003-4864-9530
S. Vig https://orcid.org/0000-0002-3477-6021
J. Martí https://orcid.org/0000-0001-5302-0660

---

[9] https://www.skao.int/en/science-users
[10] https://ngvla.nrao.edu/